\documentclass[12pt]{article}
\usepackage{fullpage,citesort,epsfig,psfrag,graphics,amsbsy,amssymb,amsmath}
\usepackage{caption}
\newcommand{\beq}{\begin{equation}}
\newcommand{\eeq}{\end{equation}}
\newcommand{\bea}{\begin{eqnarray}}
\newcommand{\eea}{\end{eqnarray}}
\newcommand{\bfs}{\boldsymbol}

\newcommand{\Tr}{{\rm Tr}}
\newcommand{\be}{\begin{equation}}
\newcommand{\ee}{\end{equation}}
\newcommand{\bq}{\begin{eqnarray}}
\newcommand{\eq}{\end{eqnarray}}
\newcommand{\ket}[1]{|#1\rangle}
\newcommand{\bra}[1]{\langle#1|}

\newcommand{\ie}{{\it i.e.\ }}
\def\math{\mathsurround=0pt }
\def\leftrightarrowfill{$\math \mathord\leftarrow \mkern-6mu
 \cleaders\hbox{$\mkern-2mu \mathord- \mkern-2mu$}\hfill
 \mkern-6mu \mathord\rightarrow$}
\def\overleftrightarrow#1{\vbox{\ialign{##\crcr
     \leftrightarrowfill\crcr\noalign{\kern-1pt\nointerlineskip}
     $\hfil\displaystyle{#1}\hfil$\crcr}}}

\newcommand{\VEV}[1]{\langle#1\rangle}

   \let\i=\iota 
\let\l=\lambda    \let\p=\pi

 \def\bd{\begin{document}} \def\ed{\end{document}}
\def\ds{\documentstyle} \let\fr=\frac \let\bl=\bigl \let\br=\bigr
\let\Br=\Bigr \let\Bl=\Bigl
\let\bm=\bibitem
\let\na=\nabla
\let\pa=\partial \let\ov=\overline
\def\ft#1#2{{\textstyle{{\scriptstyle #1}\over {\scriptstyle #2}}}}
\def\fft#1#2{{#1 \over #2}}
\def\vp{\varphi}
\def\sst#1{{\scriptscriptstyle #1}}
\def\oneone{\rlap 1\mkern4mu{\rm l}}
\def\td{\tilde}
\def\wtd{\widetilde}
\def\dalemb#1#2{{\vbox{\hrule height .#2pt
        \hbox{\vrule width.#2pt height#1pt \kern#1pt
                \vrule width.#2pt}
        \hrule height.#2pt}}}
\def\square{\mathord{\dalemb{6.8}{7}\hbox{\hskip1pt}}}
\def\wtd{\widetilde}
\def\R{\rlap{\rm I}\mkern3mu{\rm R}}
\def\im{{\rm i}}
\def\tilg{\tilde{g}}
\def\tilF{\tilde{F}}
\def\tilA{\tilde{A}}
\def\varf{\varphi}
\def\tilf{\tilde{\phi}}
\def\tilh{\tilde{h}}
\def\rme{{\rm e}}
\def\ep{\epsilon}
\def\0{{(0)}}
\def\9{{(9)}}
\def\8{{(8)}}
\def\7{{(7)}}
\def\6{{(6)}}
\def\5{{(5)}}
\def\4{{(4)}}
\def\3{{(3)}}
\def\2{{(2)}}
\def\1{{(1)}}
\newcommand{\trace}{{\rm Tr}}
\newcommand{\ub}{\overline{U}}
\newcommand{\vb}{\overline{V}}
\newcommand{\uh}{\widehat{U}}
\newcommand{\vh}{\widehat{V}}
\newcommand{\ubh}{\overline{\widehat{U}}}
\newcommand{\vbh}{\overline{\widehat{V}}}
\newcommand{\lb}{\bar{\l}}
\newcommand{\Fb}{\overline{F}}
\newcommand{\Fh}{\widehat{F}}
\newcommand{\Fbh}{\overline{\widehat{F}}}
\newcommand{\Ab}{\overline{A}}
\newcommand{\Ah}{\widehat{A}}
\newcommand{\Abh}{\overline{\widehat{A}}}
\newcommand{\Gb}{\overline{G}}
\newcommand{\Gh}{\widehat{G}}
\newcommand{\Gbh}{\overline{\widehat{G}}}
\newcommand{\Pb}{\overline{P}}
\newcommand{\Ph}{\widehat{P}}
\newcommand{\Pbh}{\overline{\widehat{P}}}
\newcommand{\Qb}{\overline{Q}}
\newcommand{\Qh}{\widehat{Q}}
\newcommand{\Qbh}{\overline{\widehat{Q}}}
\newcommand{\Bb}{\overline{B}}
\newcommand{\Bh}{\widehat{B}}
\newcommand{\Bbh}{\overline{\widehat{B}}}
\newcommand{\fhns}{\hat{F}^{\rm (NS)}}
\newcommand{\fhrr}{\hat{F}^{\rm (RR)}}
\newcommand{\ahns}{\hat{A}^{\rm (NS)}}
\newcommand{\ahrr}{\hat{A}^{\rm (RR)}}
\newcommand{\hhrr}{\hat{H}^{\rm (RR)}}
\newcommand{\hchi}{\hat{\chi}}
\newcommand{\hphi}{\hat{\phi}}
\newcommand{\htau}{\hat{\tau}}
\newcommand{\cG}{{\cal G}}
\newcommand{\cGb}{\overline{{\cal G}}}
\newcommand{\cH}{{\cal H}}
\newcommand{\cP}{{\cal P}}
\newcommand{\cPb}{\overline{{\cal P}}}
\newcommand{\cQ}{{\cal Q}}
\newcommand{\cQb}{\overline{{\cal Q}}}
\newcommand{\cM}{{\cal M}}
\newcommand{\cN}{{\cal N}}
\newcommand{\cO}{{\cal O}}
\newcommand{\cD}{{\cal D}}
\newcommand{\cL}{{\cal L}}

\newcommand{\vpp}{\mbox{$\langle{\scriptstyle++}\rangle$}}
\newcommand{\vmp}{\mbox{$\langle{\scriptstyle-+}\rangle$}}
\newcommand{\vppp}{\mbox{$\langle{\scriptstyle+++}\rangle$}}
\newcommand{\vmpp}{\mbox{$\langle{\scriptstyle-++}\rangle$}}
\newcommand{\vpmp}{\mbox{$\langle{\scriptstyle+-+}\rangle$}}

\def\narrowtext{}
\def\ie{{\it i.e.\ }}
\def\m@th{\mathsurround=0pt }
\def\leftrightarrowfill{$\m@th \mathord\leftarrow \mkern-6mu
 \cleaders\hbox{$\mkern-2mu \mathord- \mkern-2mu$}\hfill
 \mkern-6mu \mathord\rightarrow$}
\def\overleftrightarrow#1{\vbox{\ialign{##\crcr
     \leftrightarrowfill\crcr\noalign{\kern-1pt\nointerlineskip}
     $\hfil\displaystyle{#1}\hfil$\crcr}}}
\def\VEV#1{\langle#1\rangle}
\def\phdag{{\phantom{\dagger}}}
\begin{document}
\renewcommand{\thefootnote}{\fnsymbol{footnote}}
\begin{titlepage}
\begin{flushright}
\phantom{-.}
\end{flushright}
\vskip 1.5cm

\begin{center}
\begin{Large}
{\bf Space from String Bits}
\end{Large}

\vskip 2.cm

{\large Charles B. Thorn\footnote{E-mail  address: thorn@phys.ufl.edu}}

\vskip 0.5cm

{\it Institute for Fundamental Theory\\
Department of Physics, University of Florida,
Gainesville, FL 32611}


\vskip 1.5cm
\end{center}

\begin{abstract}
\noindent We develop superstring bit models, in which the lightcone
transverse coordinates in $D$ spacetime dimensions are
replaced with $d=D-2$ double-valued ``flavor''  indices $x^k\to f_k=1,2$;
$k=2,\ldots,d+1$. In such models the
string bits have no space to move. Letting each string bit be
an adjoint of a ``color'' group $U(N)$, we then analyze the
physics of 't Hooft's limit $N\to\infty$, in which closed
chains of many string bits behave like free lightcone IIB superstrings
with $d$ compact coordinate bosonic worldsheet fields $x^k$, 
and $s$ pairs of Grassmann fermionic fields $\theta_{L,R}^a$, $a=1,\ldots, s$.
The coordinates $x^k$ emerge because, on the long chains, flavor
fluctuations enjoy the dynamics of $d$ anisotropic
Heisenberg spin chains. It is well-known that the low energy
excitations of a many-spin Heisenberg chain are identical to those
of a string worldsheet coordinate compactified on a circle of radius $R_k$, 
which is
related to the anisotropy parameter $-1\leq\Delta_k\leq1$ of the corresponding
Heisenberg system. Furthermore there is a limit of this
parameter, $\Delta_k\to\pm1$, in 
which $R_k\to\infty$. 
As noted in earlier work [Phys.Rev.D{\bf 89}(2014)105002], 
these multi-string-bit 
chains are strictly stable at $N=\infty$ when $d<s$ and only marginally 
stable when 
$d=s$. (Poincar\'e supersymmetry requires $d=s=8$, which is on
the boundary between stability and instability.)
\end{abstract}
\end{titlepage}
\section{Introduction}
The idea, that string bits might provide the fundamental
constituents of string, was proposed over two decades ago 
\cite{thornsakh}, with the implications of supersymmetry for string bit models
developed and explored in \cite{bergmantsubit}.
As initially envisioned, string bits were point particles moving
about in the transverse space ${x}^k$, $k=2,\ldots,D-1$ of lightcone 
coordinates, $x^\pm=(x^0\pm x^1)/\sqrt{2}$ enjoying
a dynamics that is Galilei invariant. This Galilei invariance
is natural to lightcone coordinates in which the momentum component $P^+
=(p^0+p^1)/\sqrt{2}$ plays the role of a variable
Newtonian mass, and the Galilei transformations act on the transverse
space $x^k\to x^k+V^kx^+$. The Newtonian mass
$m$ of each string bit is fixed, but then Galilei invariance
ensures that the Newtonian mass of a bound state of $M$ string bits is $Mm$.
If bound states can form with any number $M$ of bits,
$Mm$ can be interpreted as the total (discretized) $P^+$ of the bound state.
For $M\to\infty$ this emergent $P^+$ can be regarded as a
continuous variable whose conjugate can be interpreted as $x^-$.
String theory emerges from these models, if the many-bit
bound states are closed linear chains of string bits (which
can be arranged in the context of the 't Hooft large
$N$ limit \cite{thooftlargen,thornfock}),  
in which the low lying excitation energies scale for $M\to\infty$ 
as $M^{-1}$. This
scaling law then leads to a Poincar\'e invariant dispersion relation
$P^-=({\bfs P}^2+\mu^2)/(2P^+)$. Since space is 3 dimensional with
the associated transverse space 2 dimensional, the string
bit model provided a concrete realization of 't Hooft's
idea that the world is like a hologram \cite{thoofthologram}.

Recently Sun and I have begun a new study of string bit models
in a more general context \cite{sunthorn}. Our main idea
is that relaxing the strict requirement, that super Poincar\'e 
invariance emerge, provides us with composite models of string without the
infrared instabilities caused by the massless graviton and 
gauge particles of superstring theory. Specifically, the 
lightcone quantized type II superstring requires a worldsheet system
of $8$ bosonic coordinate fields $x^k$ and $8$ pairs of
fermionic Grassmann fields $\theta^a_{L,R}$. We proposed 
studying general string bit models in which $d$ bosonic and $s$
pairs of fermionic worldsheet fields emerge, provided
that $d<s$. As long as $d<s$ the 
emergent closed string ground state has positive
mass squared ($2P^+P^--{\bfs p}^2>0$), implying that the lowest energy
closed string chain has its number of bits $M\to\infty$, i.e.
it behaves as a continuous string. The superstring,
with $d=s=8$, is only marginally stable, and when
$d>s$ long closed chains are unstable and will not form. 
As a case in point we began studying 
the simplest stable case in which the emergent string has
$d=0$ and $s=1$ with further analysis of this model to
be given in \cite{sun}.  

In this paper I would like to extend the work of \cite{sunthorn}
in a different direction by developing some of the string
bit models, proposed in that work,
which lead to composite string with general
$d<s$. The models with $d=0$ and general $s>0$ can be
obtained from \cite{bergmantsubit} by simply discarding
the dependence on transverse coordinates. Then each string
bit is an adjoint in $U(N)$ color and has $2^s$ spin
states. Half of these are bosons and half fermions.
Of course, for $s=1$ the model reduces to that studied in \cite{sunthorn},
with one boson and one fermion.
To achieve $d>0$ we could simply follow \cite{bergmantsubit} and
restore the dependence on transverse coordinates. Instead,
as already suggested in $\cite{sunthorn}$, we let each
string bit have $2^d$ ``flavor states''.
Then all together, each string bit will have $2^s2^dN^2$
internal states. It is only when these bits form long closed chains
that fluctuations among the internal bit states begin to behave like
bosonic and fermionic coordinates. Then we can say that space
has effectively emerged from string bit dynamics--hence
the title chosen for this paper.

Although string theory and the dual resonance models, which
led to its discovery, were initially developed as 
models of extended objects moving in space, it has been
understood, almost from the beginning, that the ``target space''
in which string ``moves'' need not be a continuous manifold.
Indeed, once one posits
the existence of a worldsheet, the target space can be
any two dimensional quantum field theory, which supports
a suitable Virasoro algebra with the appropriate central
charge (conformal field theories). The original
description of the evolving string as a mapping $x^\mu(\sigma,\tau)$ from a
worldsheet parameterized by $\sigma,\tau$ to space time $x^\mu$
could in the extreme be replaced by a mapping to $x^\pm(\sigma,\tau)$ 
and the values of a bunch of fermion fields $\psi_a(\sigma,\tau)$ on the 
worldsheet\footnote{By retaining $x^+$ as a 
continuous one dimensional manifold one keeps conventional
quantum dynamics, which requires
the notion of a Hamiltonian $P^-$. But this does not seem to be 
absolutely mandatory, provided that one is prepared to replace 
quantum mechanics with something else \cite{thooftfound}.}.
By bosonizing some of these fermion worldsheet fields, one can
regain (compact) coordinates. Thus string theory automatically 
provides a first step in understanding the
concept of space as an emergent phenomenon. 

The goal of string bit models is to understand the worldsheet itself
as an emergent phenomenon. In this paper,
we shall focus on models which generate a worldsheet in 
lightcone parameterization in which $x^+=\tau$ and 
$\sigma$ is chosen so the density of $P^+$
is unity, $0<\sigma<P^+$ \cite{goddardgrt}. 
To motivate them let's ``deconstruct''
the lightcone worldsheet path integral for a free closed string. 
First, the path integral
is defined on a $2$ dimensional lattice \cite{gilest},
which discretizes $\tau=x^+\to kb$ and
$\sigma\to lm$. Then we pass to Hamiltonian quantum mechanics 
by sending $b\to0$. This leaves us with a system of point
particles, each carrying a single unit of $P^+=m$, ordered on a closed 
chain enjoying nearest neighbor interactions.
Finally we embrace these string bits as fundamental
degrees of freedom which are not {\it a priori} confined
to closed chains. 
Instead of describing the  string bits by their
trajectories  ${\bfs x}_l(\tau)$, we introduce a string bit
annihilation operator $a({\bfs x})$ and an empty state $\ket{0}$
\cite{thornsakh}.
A superstring bit \cite{bergmantsubit} can be either a fermion or boson. 
And as already mentioned, the transverse space label ${\bfs x}$
can be replaced by discrete internal symmetry labels.
After this, as far as the string bits are concerned, space is
literally nonexistent. 

The emergence of the concept of space in these models depends on a 
remarkable confluence
of circumstances involving string bit dynamics. They should be such that
string bits organize themselves into closed many bit chains.
The lowest energy chains must have either an infinite or at least
an extremely large number of bits $M\gg1$. Once chains of
string bits form,
the low-lying energy excitations of a chain will be ``spin waves'' involving
fluctuations of the internal string bit states 
(including fluctuations of statistics!).
For large $M$ these low excitation energies will naturally scale as $M^{-1}$.
This leads to the interpretation of the chain energy  eigenstate as 
a particle moving in one space dimension, with lightcone
dispersion relation $P^-=\mu^2/2P^+$. The spectrum of particle masses
$\mu^2$ depends on the nature of the internal fluctuation waves
allowed by the dynamics.
Thus the Hamiltonian giving string bit dynamics is interpreted as $P^-$ and
the bit number as $P^+=Mm$. The longitudinal dimension $x^-$
therefore emerges as the conjugate to $P^+$.

A straightforward way to set up the string bit dynamics
to favor chain formation is to exploit the 't Hooft large $N$
expansion \cite{thooftlargen}. This is done \cite{thornfock,thornsakh}
by letting the annihilation operator 
for a string bit be an $N\times N$ matrix $(a_K)_\alpha^{\ \beta}$,
and choosing the Hamiltonian as a sum of terms with the structure
\bea
\frac{2}{N}\Tr a_K^\dagger a_L^\dagger a_I a_J
\eea
Then when $N\to\infty$ the Hamiltonian connects single trace states to 
single trace states with only nearest neighbor interactions,
which sets up a one dimensional spin chain problem. In
\cite{sunthorn} we observed that long chains will be energetically
favored if the number of statistics fluctuating waves $s$ exceeds the
number of statistics nonfluctuating waves $d$. We then studied in great
detail a model where there was precisely one of the former and none
of the latter. For supersymmetry these
two types of waves are equal in number, so the string bit model
underlying superstring is on the boundary between stability and instability. 

In the next Section 2, we present the string bit models studied in
this paper. Then in Section 3 we explain how long closed chains form
dynamically and thereby convert fluctuating internal spin states
to Grassmann worldsheet fields. In Section 4 we discuss
the Heisenberg spin chain. We present the Bethe
ansatz \cite{bethe} for its energy eigenstates 
for general anisotropy parameter $\Delta$. For $\Delta=0$ it
is easy to read off the energy spectrum for $M$ spins and evaluate its large
$M$ behavior. For $\Delta\neq0$ Yang and Yang \cite{yangyang}
have analyzed the energy spectrum for general $\Delta$ and $M\gg1$.
We review their analysis in an Appendix, with particular attention 
to the aspects relevant to the present paper. 
Then in Section 5 we explain how the formation
of long chains converts the internal flavor states
to flavor waves described by the Heisenberg Hamiltonian. 
Section 6 closes the paper with a preliminary discussion of
string interactions together with some concluding remarks.
 
\section{Superstring bit models}
The type IIB superstring theory \cite{gso,greenschwarz}
quantized on the lightcone \cite{goddardgrt}
is based on
a worldsheet system with 8 transverse coordinates $x^k(\sigma,\tau)$,
and 8 left moving and 8 right moving Grassmann variables 
$\theta^a_{L,R}(\sigma,\tau)$. We contemplate a more general 
worldsheet system with $d$ coordinates and $s$ pairs of Grassmann
variables. The values $d=s=8$ are necessary for Poincar\'e
supersymmetry, but the general case, which lacks these symmetries,
is a perfectly sensible dynamical system worthy of study in its own
right. Even though full Poincar\'e invariance is
lost when $s\neq 8$ and/or $d\neq8$, the lightcone dynamics still
naturally implements a relativistic energy momentum dispersion
law $P^-=({\bfs p}^2+\mu^2)/2P^+$ in the 't Hooft limit, so the concept of 
particle mass is retained at least in the limit $N\to\infty$.
In particular, it was emphasized in \cite{sunthorn} that
when $s>d$ there is a gap (i.e. the lowest $\mu^2>0$)  
in the mass spectrum of the
string system, which tames infrared divergences. This
gap vanishes for $d=s$ and is tachyonic $\mu^2<0$, leading
to instabilities, when $d>s$. Because the closed chains are
noninteracting when $N=\infty$, these instabilities are
problematic only at finite $N$. 

As mentioned in the introduction we set up superstring bit dynamics
in the standard second-quantized formalism, with creation and
annihilation operators for a string bit.
In general, a superstring bit annihilation operator is
an $N\times N$ matrix denoted by 
\bea
\left(\phi_{[a_1\cdots a_n]}^{f_1\cdots f_d}\right)_{\alpha}^{\ \beta},\qquad
n=0,\ldots,s;\quad a_j=1,\ldots,s;\quad \alpha,\beta=1,\ldots,N
\eea
The $f_j$ are ``flavor'' indices, describing the degrees of
freedom responsible for the emergence of transverse space. The $a_k$ 
are spinor indices, with the square brackets enclosing them
reminding us that $\phi$ is completely antisymmetric under permutations
of them. Also $\phi$ will be bosonic (fermionic) 
if the number of spinor indices $n$ is even (odd). 
We will denote the corresponding creation operator 
by ${\bar\phi}_\alpha^{\ \beta}
\equiv (\phi_\beta^{\ \alpha})^\dagger$.
In the simplest realization of
$d$ space coordinates proposed in \cite{sunthorn}, it is enough that each 
$f_j=1,2$. In
that model for the superstring, the Hamiltonian is
\bea
H&=&H_F+H_S
\eea
where the flavor dynamics is given by
\bea
H_F&=&\frac{2}{N}\sum_{n=0}^s\sum_{k=0}^s\frac{1}{n!k!}
\Tr{\bar\phi}^E_{a_1\cdots a_n}{\bar\phi}^F_{b_1\cdots b_k}
{\phi}^G_{b_1\cdots b_k}{\phi}^H_{a_1\cdots a_n}V_{EFGH}
\eea
In this formula each capital superscript represents the 
collection of the individual flavor indices $F=\{f_j\}$.
The spinor dynamics is described by
\bea
H_S&=& H_1+H_2+H_3+H_4+H_5
\eea
where the $H_i$ are:
\bea
H_1&=&\frac{2}{N}\sum_{n=0}^s\sum_{k=0}^s\frac{s-2n}{n!k!}
\Tr{\bar\phi}^F_{a_1\cdots a_n}{\bar\phi}^G_{b_1\cdots b_k}
{\phi}^G_{b_1\cdots b_k}{\phi}^F_{a_1\cdots a_n}\\
H_2&=&\frac{2}{N}\sum_{n=0}^{s-1}\sum_{k=0}^{s-1}\frac{(-)^k}{n!k!}
\Tr{\bar\phi}^F_{a_1\cdots a_n}{\bar\phi}^G_{bb_1\cdots b_k}
{\phi}^G_{b_1\cdots b_k}{\phi}^F_{ba_1\cdots a_n}\\
H_3&=&\frac{2}{N}\sum_{n=0}^{s-1}\sum_{k=0}^{s-1}\frac{(-)^k}{n!k!}
\Tr{\bar\phi}^F_{ba_1\cdots a_n}{\bar\phi}^G_{b_1\cdots b_k}
{\phi}^G_{bb_1\cdots b_k}{\phi}^F_{a_1\cdots a_n}\\
H_4&=&\frac{2i}{N}\sum_{n=0}^{s-1}\sum_{k=0}^{s-1}\frac{(-)^k}{n!k!}
\Tr{\bar\phi}^F_{a_1\cdots a_n}{\bar\phi}^G_{b_1\cdots b_k}
{\phi}^G_{bb_1\cdots b_k}{\phi}^F_{ba_1\cdots a_n}\\
H_5&=&-\frac{2i}{N}\sum_{n=0}^{s-1}\sum_{k=0}^{s-1}\frac{(-)^k}{n!k!}
\Tr{\bar\phi}^F_{ba_1\cdots a_n}{\bar\phi}^G_{bb_1\cdots b_k}
{\phi}^G_{b_1\cdots b_k}{\phi}^F_{a_1\cdots a_n}
\eea
The formula for $H_S$ can be inferred from the superstring bit
Hamiltonian proposed in \cite{bergmantsubit} by discarding all contributions
from the transverse coordinates.
The structure of $H$ is designed so that the action of $H$ on
single trace states goes, in the 't Hooft limit $N\to\infty$ 
\cite{thooftlargen,thornfock}, to the action of 
a discretized version of the first quantized Hamiltonian
for the worldsheet fields of the type IIB superstring. 

To explain this we introduce $s$ Grassmann variables
$\theta^a$ and construct the super creation operators
\bea
\psi^F(\theta)&=&\sum_{k=0}^s
\frac{1}{k!}{\bar\phi}^F_{c_1\cdots c_k}\theta^{c_1}
\cdots\theta^{c_k}\eea
so that a general single trace state can be constructed from
\bea
\ket{\theta_1F_1,\cdots,\theta_MF_M}&=&\Tr\psi^{F_1}(\theta_1)\cdots
\psi^{F_M}(\theta_M)\ket{0}.
\eea
We note that the cyclic property of the trace implies the cyclic
symmetry condition
\bea
\ket{\theta_1F_1,\cdots,\theta_MF_M}&=&
\ket{\theta_2F_2,\cdots,\theta_MF_M,\theta_1F_1}
\label{cyclicket}
\eea
Then we evaluate
\bea
H_F\ket{\theta_1F_1,\cdots,\theta_MF_M}&\hskip-9pt=&\hskip-9pt
2\hskip-4pt\sum_{k=1}^M\ket{\theta_1F_1,\cdots,\theta_kG,\theta_{k+1}G^\prime,
\cdots,\theta_MF_M}V_{GG^\prime F_{k+1}F_{k}}+\mathcal{O}(N^{-1})
\eea
\bea
H_1\ket{\theta_1F_1,\cdots,\theta_MF_M}&=&2\sum_{k=1}^M\left(s-2\theta_k^a
\frac{d}{d\theta_k^a}\right)
\ket{\theta_1F_1,\cdots,\theta_MF_M}+\mathcal{O}(N^{-1})
\eea
\bea
H_2\ket{\theta_1F_1,\cdots,\theta_MF_M}&=&2\sum_{k=1}^M
\theta_k^a\frac{d}{d\theta_{k+1}^a}\ket{\theta_1F_1,\cdots,
\theta_MF_M}+\mathcal{O}(N^{-1})
\\
H_3\ket{\theta_1F_1,\cdots,\theta_MF_M}&=&2\sum_{k=1}^M
\theta_{k+1}^a\frac{d}{d\theta_{k}^a}
\ket{\theta_1F_1,\cdots,\theta_MF_M}+\mathcal{O}(N^{-1})
\eea
\bea
H_4\ket{\theta_1F_1,\cdots,\theta_MF_M}&=&-2i\sum_{k=1}^M
\theta_k^a\theta^a_{k+1}
\ket{\theta_1F_1,\cdots,\theta_MF_M}+\mathcal{O}(N^{-1})+\mathcal{O}(N^{-1})\\
H_5\ket{\theta_1F_1,\cdots,\theta_MF_M}&=&-2i\sum_{k=1}^M
\frac{d}{d\theta_k^a}\frac{d}{d\theta^a_{k+1}}
\ket{\theta_1F_1,\cdots,\theta_MF_M}+\mathcal{O}(N^{-1})
\eea
To formulate the energy spectrum problem at $N=\infty$ in the first
quantized language we express the sought eigenstate as
\bea
\ket{E}&=&\int d^s\theta_1\cdots d^s\theta_M\ket{\theta_1F_1,
\cdots,\theta_MF_M}\Psi(\theta_1,\ldots,\theta_M)U_{F_1\cdots F_M}.
\eea
Because of the cyclic property (\ref{cyclicket}), we can, without loss 
of generality, require the wave function to satisfy
\bea
\Psi(\theta_1,\ldots,\theta_M)U_{F_1\cdots F_M}&=&(-)^{s(M-1)}
\Psi(\theta_2,\ldots,\theta_M,\theta_1)U_{F_2\cdots F_MF_1},
\label{cyclicwf}
\eea
where the sign out front arises from reordering the $d^s\theta_k$'s.
The wave functions must be anticyclic if both $s$ and $M-1$ are odd,
and cyclic otherwise.

We now apply $H$ to $\ket{E}$ and write out the eigenvalue condition:
\bea
E\ket{E}&=&H\ket{E}=\int d^s\theta_1\cdots d^s\theta_M\ket{\theta_1F_1,\cdots,
\theta_MF_M}h\left(\Psi(\theta_1,\ldots,\theta_M)U_{F_1\cdots F_M}\right)
\eea
for $N=\infty$.
An integration by parts in the $\theta$ variables is done in the last step,
after which one finds $h=h_S+h_F$ with
\bea
h_S=2\sum_{k=1}^M\left[-i\theta_k^a\theta^a_{k+1}
-i\frac{d}{d\theta_k^a}\frac{d}{d\theta^a_{k+1}}
-\theta_k^a\frac{d}{d\theta_{k+1}^a}
-\theta_{k+1}^a\frac{d}{d\theta_{k}^a}-s+2\theta_k^a
\frac{d}{d\theta_k^a}\right],
\eea
and $h_F$ acts as a matrix 
\bea
\bra{G_1\cdots G_M}h_F\ket{F_1\cdots F_M}=\sum_{k=1}^M \delta_{G_1}^{F_1}
\cdots\delta_{G_{k-1}}^{F_{k-1}}V_{G_kG_{k+1}F_{k+1}F_k}
\delta_{G_{k+2}}^{F_{k+2}}\cdots\delta_{G_M}^{F_M}
\eea
In other words $h_F$ is a sum of $M$ terms, where the $k$th term acts as
the identity on all the indices of $U$ except for $F_k$ and $F_{k+1}$,
and acts on these two as a matrix. As a useful shorthand in the 
first quantized dynamics we can write 
\bea
h_F&=&\sum_{k=1}^M V^k\\
\bra{G_1\cdots G_M}V^k\ket{F_1\cdots F_M}
&=&\delta_{G_1}^{F_1}
\cdots\delta_{G_{k-1}}^{F_{k-1}}V_{G_kG_{k+1}F_{k+1}F_k}
\delta_{G_{k+2}}^{F_{k+2}}\cdots\delta_{G_M}^{F_M}
\eea
This nearest neighbor action presents the flavor
dynamics of the string bit system as a generalized 
one dimensional chain.
Since the low energy states of a Heisenberg spin chain with $M$ spins behave 
at large $M$ like those of a compactified coordinate, we can
choose $h_F$ to be the sum of $d$ independent spin
chain Hamiltonians:
\bea
h_F=\sum_{l=1}^d C_l\sum_{k=1}^M
\left[(\sigma_l)_k^1(\sigma_l)^1_{k+1}+(\sigma_l)_k^2(\sigma_l)^2_{k+1}
+\Delta_l(\sigma_l)_k^3(\sigma_l)^3_{k+1}\right]
\eea
Here, for each $l,k$, $(\sigma_l)_k^{1,2,3}$ are the two by two Pauli 
spin matrices
\bea
\sigma^1&=&\begin{pmatrix}0&1\\1&0\end{pmatrix},\qquad
\sigma^2=\begin{pmatrix}0&-i\\i&0\end{pmatrix},\qquad
\sigma^3=\begin{pmatrix}1&0\\0&-1\end{pmatrix}
\eea
The first quantized Hamiltonian $h_F$ is the sum of $d$ 
commuting operators, each the Hamiltonian of a Heisenberg spin system.
Studies by Bethe \cite{bethe} and Yang and Yang \cite{yangyang},
which we review in section \ref{heischain} and the appendix, enable the exact
calculation of the low lying energy spectrum of the 
spin chain when $M\to\infty$. As explained in \cite{gilesmt} these energy 
eigenstates
are identical to those of a spatial coordinate compactified
on a circle. In addition to string vibrational modes, there are
two zero modes, one corresponding to the string momentum and the other
to winding the string around the compactification circle. The
radius of the circle is related to the parameter $\Delta_l$,
so each dimension can be compactified on a different sized circle.
The infinite radius limit, corresponding to a noncompact
spatial dimension, is reached by the limit $\Delta_l\to1$, if
$C_l<0$ and $\Delta_l\to-1$ if $C_l>0$.  

Requiring this outcome determines the coefficients $V_{GG^\prime F^\prime F}$
of the string bit Hamiltonian. Each capital index is a string
of two-valued indices $F=\{f_1\cdots f_d\}$ and similarly for
$G,F^\prime,G^\prime$. Then
\bea
V_{GG^\prime F^\prime F}&=&\sum_{l=1}^d C_l
\left[\sigma^1_{g_lf_l}\sigma^1_{g_l^\prime f_l^\prime}
+\sigma^2_{g_lf_l}\sigma^2_{g_l^\prime f_l^\prime}
+\Delta_l\sigma^3_{g_lf_l}\sigma^3_{g_l^\prime f_l^\prime}
\right]\prod_{k\neq l}\left(\delta_{g_kf_k}\delta_{g_k^\prime f_k^\prime}
\right).
\eea
When there is no danger of confusion, we will suppress the indices on the right
side of this equation which we can write:
\bea
V_{GG^\prime F^\prime F}&\to&\sum_{l=1}^d C_l
\left[\sigma^1_{l}\sigma^{1\prime}_{l}
+\sigma^2_{l}\sigma^{2\prime}_{l}
+\Delta_l\sigma^3_{l}\sigma^{3\prime}_{l}
\right].
\eea
In this notation the $\sigma^{1,2,3}_l,\sigma^{\prime1,2,3}_l $ 
commute with the
$\sigma^{1,2,3}_{l^\prime},\sigma^{\prime1,2,3}_{l^\prime}$ 
when $l\neq l^\prime$ and the $\sigma^{1,2,3}_l$
commute with the $\sigma^{\prime1,2,3}_l $.

\section{Emergence of Grassmann and longitudinal space}
At finite $N$ the string bit system has a finite number of degrees of freedom:
for our example this number is $2^{s+d}N^2$. In the large $N$ limit
the low energy eigenstates with a large bit number $M$ 
show an energy excitation of order $1/M$. Recalling that
the lightcone mass shell condition for a particle is
$P^-=(m^2+{\bfs p}^2)/2P^+$, we seek to interpret $M$ as $P^+$
and $H$ as $P^-$. The eigenvalues of $h_S$ at fixed $M$ are
\cite{bergmantsubit,sunthorn} 
\bea
E_S&=&E_S^{\rm min}+8\sum_{n=1}^{M-1}\eta_n\sin\frac{n\pi}{M},\qquad 
\eta_n=0,1,\ldots,s \\
E_S^{\rm min}&=& -4s\cot\frac{\pi}{2M}\sim -\frac{s8M}{\pi}+\frac{2\pi s}{3M}
\eea
corresponding to states built from fermionic creation operators
$B^{a\dagger}_n$, $n=0,\ldots,M-1$, applied to a ground state $\ket{0}$.
Here the modes $n$ and $M-n$ have the same frequency. Excitations
of order $1/M$ arise when $M\to\infty$ with either $n/M\ll1$ or $(M-n)/M\ll1$.
We can call the first case left moving modes and the second
right moving modes.
When $s$ is even, the cyclic symmetry constraint amounts to the
requirement that $N_L=N_R$ where $N_L$, $N_R$ are the total mode numbers
of left moving right moving modes respectively. The zero mode $B_0$
converts boson states to fermion states
and vice versa. Its existence establishes that the number of bosonic
states is the same as the number of fermionic states.

When $s=8$ this excitation spectrum is precisely that of the left and right
moving Grassmann worldsheet fields $\theta^a_L,\theta^a_R$ of the 
Green-Schwarz formulation of the type IIB superstring. 
This identification fixes the scale of the energy in terms of 
$m$ the unit of $P^+\equiv Mm$
and the rest tension of the string $T_0=1/(2\pi\alpha^\prime)$. Each worldsheet
coordinate field should contribute $-\pi T_0/(6P^+)$ to the closed string
ground state $P^-$. To see this just consider the bosonic string which
has 24 such coordinate worldsheet fields. We know that the ground state
mass squared of the bosonic closed string is $2P^+P^-=-4/\alpha^\prime
=-8\pi T_0$. For the superstring each left-right pair
of Grassmann fields contribute
just the negative of this, namely $\pi T_0/(6P^+)$. Remembering that
$P^+=Mm$, it follows that $P_S^-=E_S T_0/(4m)$, or for operators
\bea
P^-&=& P^-_S + P^-_F\\
P^-_S&=&\frac{T_0}{4m}(H_1+H_2+H_3+H_4+H_5)
\eea
and $P^-_F$ will be determined after the analysis of the next section. 
\section{Heisenberg Spin Chain}
\label{heischain}
The conventional Hamiltonian for the spin chain is usually
defined as
\bea
H_{hei}=-\sum_{k=1}^M
\left(\sigma_k^1\sigma^1_{k+1}+\sigma_k^2\sigma^2_{k+1}
+\Delta\sigma_k^3\sigma^3_{k+1}\right)
\eea
The application to string bit
dynamics requires periodic boundary conditions, meaning $\sigma_{M+1}
\equiv\sigma_1$, which we henceforth assume.
It is well-known that critical behavior is present for the range 
$-1\leq\Delta\leq1$. The
minus sign out front nominally favors spin alignment, but for 
$\Delta$ in this range the lowest energy states actually have 
charge (spin) $Q=\sum_k\sigma_k^3=0$. The solution for the eigenvalues
of $H$ is given by the Bethe ansatz \cite{bethe}, in which eigenstates are
sought as spin waves of overturned spins relative to the
state $\ket{0}$, in which $\sigma_k^3=+1$ for all $k$. Then a state
with $q$ overturned spins is denoted $\ket{i_1,\cdots,i_q}$
where the $i_k$ give the locations of the overturned spins. 
Then the Bethe
ansatz is
\bea
\ket{E}&=&\sum_P A_P\ket{i_1,\cdots,i_q}e^{i\sum_ki_kp_{{}_{P_k}}}\\
H\ket{0}&=&-M\Delta\ket{0},\qquad E=-M\Delta+4\sum_{k=1}^q(\Delta-\cos p_k).
\eea
Here, the sum over $P$ is the sum of all permutations of $12\cdots q$. 
Without loss of generality, we may restrict $-\pi\leq p_k\leq\pi$.
The ansatz is an eigenstate of $H$, provided that the $p_k$,
which must all be distinct, satisfy
\bea
p_k&=&\frac{2\pi I_k}{M}-\frac{1}{M}\sum_{j=1}^q\theta(p_k,p_j)
\label{bethekernel}\\
\theta(p_k,p_j)&\equiv&2\arctan\frac{\Delta\sin((p_k-p_j)/2))}
{\cos((p_k+p_j)/2)-\Delta\cos((p_k-p_j)/2)}
\eea
Periodic boundary conditions require that the $I_k$ are
integers when $q$ is odd, or that they are half odd integers when
$q$ is even. We can identify two conserved quantities that
help characterize the different eigenstates especially at 
large $M$. One is the total
charge $Q=M-2q$ for the state with $q$ overturned spins.
The other is the total momentum of the overturned spins 
\bea
P&=&\sum_k p_k=\frac{2\pi}{M}\sum_k I_k.
\eea
The last equality follows from the antisymmetry of $\theta(k,k^\prime)=
-\theta(k^\prime,k)$.
Inspection of the form of the energy eigenvalue shows that the
energy is minimized (maximized) for the maximum value of $q$
for which all of the overturned spins satisfy $\Delta-\cos p_k<0$ ($>0$).
\subsection{Energy Analysis for $\Delta=0$.}
The equation (\ref{bethekernel}) is formidable but Yang and Yang
have successfully analyzed it for large $M$ \cite{yangyang}.
We review their analysis in the appendix. Here we discuss the
case $\Delta=0$ for which $\theta=0$, and hence $p_k=2\pi I_k/M$
exactly for all $M$.
Focusing first on lowest energy states, we are
interested in all the $p_k$ in the range $-\pi/2<p_k<\pi/2$. When
$q$ is even (odd) the $p_k$ are of the form $\pi(2n_k+1)/M$
($2\pi n_k/M$). For fixed $M,q$ the energy is minimized when the
$n_k$ are consecutive integers as symmetrical about $0$
as possible. So let the $n_k$ range from $n_1$ to $n_2$ spaced
by integers. Then $q=n_2-n_1+1$, and
\bea
E&=&\begin{cases}
{\displaystyle-4\sum_{k=n_1}^{n_2}\cos\frac{2\pi k}{M}
=-2\frac{\sin[\pi(2n_2+1)/M]
-\sin[\pi(2n_1-1)/M]}{\sin[\pi/M]}},& q\quad{\rm odd}\\
{\displaystyle-4\sum_{k=n_1}^{n_2}\cos\frac{\pi(2k+1)}{M}
=-2\frac{\sin[2\pi(n_2+1)/M]
-\sin[2\pi n_1/M]}{\sin[\pi/M]}},& q\quad{\rm even}\end{cases}
\eea
For the given $M$, find an integer $r$ such that $M-r$ is divisible
by $4$. We may take $r$ from the set $\{-1,0,1,2\}$.
Then define $l,k$ by $q=(M-r)/2+l$ and $n_1=-(M-r)/4+k$.
It follows that $Q=r-2l$, $n_2=l+k-1+(M-r)/4$.
Note that $q$ is even (odd) if and only if $l$ is even (odd).
Plugging these expressions into the energy formulas,
\bea
E&=&\begin{cases}
{\displaystyle-2\frac{\cos[\pi(-Q+2k-1+r/2)/M]
+\cos[\pi(2k-1+r/2)/M]}{\sin[\pi/M]}},& l\ {\rm odd}\\
\\
{\displaystyle-2\frac{\cos[\pi(-Q+2k+r/2)/M]
+\cos[\pi(2k+r/2)/M]}{\sin[\pi/M]}},& l\ {\rm even}\end{cases}
\eea
In this form we can take the long chain limit $M\to\infty$ with
$l,k$ fixed, and identify the low-lying excitations.
\bea
E&\sim&\begin{cases}
{\displaystyle4\left[-\frac{M}{\pi}-\frac{\p}{6M}
+\frac{\pi}{M}\left(\frac{Q^2}{8}+2\left\{k+\frac{l-1}{2}\right\}^2\right)
\right]},& l\ {\rm odd}\\
\\
{\displaystyle4\left[-\frac{M}{\pi}-\frac{\p}{6M}
+\frac{\pi}{M}\left(\frac{Q^2}{8}+2\left\{k+\frac{l}{2}\right\}^2\right)
\right]},& l\ {\rm even}\end{cases}
\eea
Note that whether $l$ is even or odd, the quantity in braces is
any integer. Also, $Q$ is an even (odd) integer if $M$ is even (odd).
This quantity has a simple interpretation in terms of the total
momentum $P$:
\bea
P&=&\begin{cases}{\displaystyle\sum_{k=n_1}^{n_2}\frac{2\pi k}{M}
=\frac{\pi}{M}(l+2k-1)
\frac{M+2l-r}{2}\sim\pi\left(k+\frac{l-1}{2}\right)}& l\ {\rm odd}\\
\\
{\displaystyle\sum_{k=n_1}^{n_2}\frac{\pi(2k+1)}{M}=\frac{\pi}{M}(l+2k)
\frac{M+2l-r}{2}\sim\pi\left(k+\frac{l}{2}\right)}& l\ {\rm even}\end{cases}
\eea
where the last forms are for $M\to\infty$. Thus we can write the
final answer 
\bea
E&\sim&
4\left[-\frac{M}{\pi}-\frac{\p}{6M}
+\frac{\pi}{M}\left(\frac{Q^2}{8}+2\frac{P^2}{\pi^2}\right)
\right]
\eea
where $Q$ and $P/\pi$ can be any pair of integers, with the evenness or 
oddness of $Q$ correlated with that of $M$.

The energy spectrum for $\Delta\neq0$ involves the full
sophistication of the Bethe ansatz. Once it is fully implemented,
one can take the large $M$ limit to find the low lying
energy spectrum of $H(\Delta)$. The analysis of \cite{yangyang},
reviewed in the appendix,
shows that the only effect of $\Delta\neq0$ is
to alter the overall constant out front and
the coefficients of $M$, $Q^2$ and $P^2$:
\bea
E&\sim&
\frac{2\pi\sin\mu}{\mu}\left[-\alpha(\mu)\frac{M}{\pi}-\frac{\p}{6M}
+\frac{\pi}{M}\left(\frac{\pi-\mu}{\pi}\frac{Q^2}{4}+\frac{\pi}{\pi-\mu}
\frac{P^2}{\pi^2}+2(N_L+N_R)\right)
\right]\label{ferroen}\\
\Delta&=&-\cos\mu
\eea
The term involving $(N_L+N_R)$ gives the spin wave excitations, 
which correspond to particle-hole configurations
of the momenta near the ``Fermi sea'' of the overturned spins. 
The $-\pi/(6M)$ term is the universal part of the zero point energy 
associated with these particle-hole excitations \cite{brinknielsen}.

\subsection{Lowest lying energies of $H^\prime=-H_{\rm Hei}$}
An interesting twist occurs if the Hamiltonian is taken to be
$-H$. Then the lowest energy states correspond to the highest energy 
eigenstates of $H$. To find these for $\Delta=0$, we now require that the $p_k$
all satisfy $\cos p_k<0$. Thus $\pi/2<p_k\leq\pi$ or
$-\pi< p_k<-\pi/2$. Adding $2\pi$ to the $p_k$ in the second
category allows the two categories to be unified to
$\pi/2<p_k<3\pi/2$. So now we define $k,l$ by $n_1=k+(M-r)/4$,
$q=l+(M-r)/2$, $n_2=k+l-1+3(M-r)/4$. Using these definitions the energy
can be written
\bea
E&=&\begin{cases}
{\displaystyle2\frac{\cos[\pi(2k+2l-1-3r/2)/M]
+\cos[\pi(2k-1-r/2)/M]}{\sin[\pi/M]}},& l\quad{\rm odd}\\
\\
{\displaystyle2\frac{\cos[\pi(2k+2l-3r/2)/M]
+\cos[\pi(2k+(M-r)/2)/M]}{\sin[\pi/M]}},& l\quad{\rm even}\end{cases}\\
E&\sim&\begin{cases}
{\displaystyle4\left[\frac{M}{\pi}+\frac{\p}{6M}
-\frac{\pi}{M}\left(\frac{Q^2}{8}+2\left\{k+\frac{l-1-r}{2}\right\}^2\right)
\right]},& l\ {\rm odd}\\
\\
{\displaystyle4\left[\frac{M}{\pi}+\frac{\p}{6M}
-\frac{\pi}{M}\left(\frac{Q^2}{8}+2\left\{k+\frac{l-r}{2}\right\}^2\right)
\right]},& l\ {\rm even}\end{cases}
\eea
where the second line shows the large $M$ behavior. In this case the
quantities in braces are integers when $r$ (and also $M$) are even
and  half odd integers when $r$ (and also $M$) are odd. In the latter
odd case, $Q$ is also odd, with the implication that when $M$ is odd
neither of the squared terms can ever be zero. 

With the $p_k$ in the range $\pi/2<p_k<3\pi/2$, as we have chosen
here, the total momentum works out to:
\bea
P&=&\begin{cases}{\displaystyle\frac{\pi}{M}(M-r+l+2k-1)
\frac{M+2l-r}{2}\sim q\pi+\pi\left(k+\frac{l-1-r}{2}\right)}& l\ {\rm odd}\\
{\displaystyle\frac{\pi}{M}(M-r+l+2k)
\frac{M+2l-r}{2}\sim q\pi
+\pi\left(k+\frac{l-r}{2}\right)}& l\ {\rm even}\end{cases}
\eea
where in the last forms, we have dropped some
terms of order $M^{-1}$. We see that the quantity $(P-q\pi)/\pi$
approaches the quantities in braces, so that we can write the
highest energies as
\bea
E&\sim&4\left[\frac{M}{\pi}+\frac{\p}{6M}
-\frac{\pi}{M}\left(\frac{Q^2}{8}+2\frac{(P-q\pi)^2}{\pi^2}\right)
\right]
\eea
As we have mentioned the high energy spectrum of $H$ becomes the
low energy spectrum of $-H$. When $M$ is even these two Hamiltonians
are similar. 
We have just found that when $M$ is odd, the energy spectrum of
$-H(0)$ is {\it not} the same as $H(0)$. This implies that the
two Hamiltonians are not similar when $M$ is odd.

For $\Delta\neq0$, the highest eigenvalues of $H$ are 
still the lowest eigenvalues of $-H_{\rm Hei}$:
\bea
H^\prime&=&-H_{\rm Hei}(\Delta)=\sum_{k=1}^M
\left(\sigma_k^1\sigma^1_{k+1}+\sigma_k^2\sigma^2_{k+1}
+\Delta\sigma_k^3\sigma^3_{k+1}\right).
\eea
When $M$ is even, the operator 
$C=\prod_{k=\rm odd}\sigma^3_k=C^{-1}=C^\dagger$
relates $H^\prime$ to the original Heisenberg Hamiltonian with
$\Delta\to-\Delta$.
\bea
H^\prime&=&CH_{\rm Hei}(-\Delta)C
\eea
It immediately follows that the lowest energy eigenvalues of $H^\prime$
are those of $H_{\rm Hei}$ with $\mu\to\pi-\mu$:
\begin{eqnarray}
E^\prime-E^\prime_0&=&\frac{2\pi\sin\mu}{\pi-\mu}
\left[-\frac{\pi}{6}+{\mu\over4}{Q^2}+{1\over\mu}{{\hat P}^2}
+{2\pi(N_R+N_L)}\right]\frac{1}{M},\qquad M\quad{\rm even}
\end{eqnarray}
Here ${\hat P}$ is the total momentum in the Bethe ansatz for
an eigenstate $\ket{\{p_k\}}$ of $H_{\rm Hei}(-\Delta)$.
The corresponding eigenstate of $H^\prime$ is $C\ket{\{p_k\}}$.
The operator $C$ multiplies each term, in the Bethe ansatz with
an odd number of overturned spins sitting on odd sites, by
$-1$. But this is equivalent to adding $\pi$ to each of the
$p_k$. Indeed, it is easy to show that if the $p_k$ satisfy the
Bethe ansatz conditions for $-\Delta$, then $p_k+\pi$ satisfy
the Bethe ansatz conditions for $\Delta$. Thus the total momentum 
in the Bethe ansatz for $H^\prime$ is $P={\hat P}+q\pi$. thus we can
write
\bea
E^\prime-E^\prime_0&=&\frac{2\pi\sin\mu}{\pi-\mu}
\left[-\frac{\pi}{6}+{\mu\over4}{Q^2}+{1\over\mu}(P-q\pi)^2
+{2\pi(N_R+N_L)}\right]\frac{1}{M}.
\label{antiferroen}\eea
We remind the reader that we have been assuming $M$ is even,
to rigorously obtain these results without additional work.
However the continuum analysis that leads directly to all
these results is quite insensitive to  the evenness
of oddness of $M$. Hence the restriction to even $M$ can
be dropped. The discrete nature of $Q$ and $P$ is sensitive
to the parity of $M$, but not to the value of $\Delta$. Thus
$Q=M-2q$ is even (odd) if $M$ is even (odd). Similarly
$P=(2\pi/M)\sum_k I_k$ assumes for all $\Delta$ the values 
it has for $\Delta=0$. In the low lying states of $H_{\rm Hei}$
these are integer multiples of $\pi$ for all $M$, even and odd.
But in the high lying states of $H_{\rm Hei}$
(low lying states of $H^\prime$), the values of $P-q\pi$ 
are integer multiples of
$\pi$ for even $M$, but half odd integer multiples of $\pi$
for odd $M$.
\section{Emergence of transverse space}
We obtain the normalization of $H_F$ for the case where the flavor
dynamics is given by a collection of Heisenberg spin chains from
the analysis of the previous section. The key is the coefficient of the
$1/M$ terms. Each transverse dimension should contribute $-\pi T_0/(6Mm)$
to this part of $P^-$. The lowest energy eigenvalue of $H_F$ is
$\sum_k|C(\mu_k)|E(\mu_k)$ where $E(\mu)$ is given by (\ref{ferroen})
if $C_k<0$ and by (\ref{antiferroen}) if $C_k>0$. Matching the $1/M$ terms
gives $C_k=-\mu T_0/(2\pi m\sin\mu)$ in the first case and
$C_k=(\pi-\mu)T_0/(2\pi m\sin\mu)$ in the second case.

To understand the interpretation of (\ref{ferroen}) or
(\ref{antiferroen}) as the excitations of transverse coordinates
we compare them to the spectrum of the lightcone string Hamiltonian
\bea
P^-=\frac{1}{2}\int_0^{P^+}d\sigma\left[{{\cal P}}^2
+T_0^2{ x}^{\prime2}\right]
\eea given by
\bea
P^-&=&\frac{\pi T_0}{P^+}\left[-\frac{1}{6}+\frac{2\pi}{L^2 T_0} n^2
+\frac{L^2 T_0}{2\pi}l^2+2(N_L+N_R)\right]
\eea
where the coordinate $x$ lives on a circle of circumference $L$.
The momentum is $p=2\pi n/L$. and $l$ is the number of times the
closed string winds around the circle. Comparison then shows that
the Heisenberg spin system describes a coordinate compactified on a circle
with circumference determined by
\bea
\frac{2\pi}{L^2 T_0}&=& \frac{\pi-\mu}{\pi}\quad {\rm or}\quad\frac{\mu}{\pi}.
\eea
The interpretation is exact in the sector with $Q$ an even integer,
which is the even $M$ subspace. Including the $Q$ odd sector adds
half integer values of $n$ to the momentum, corresponding to
antiperiodic boundary conditions. For $C_k<0$ the values of $l$
remain integral, but for $C_k>0$ the $l$ associated with half odd $n$
are also half odd. This has the consequence that the decompactification
limit sends the $M$ odd chains to infinite energy in the second case. 
\section{Conclusion}
In this paper we have presented a class of stable string bit models in
which space is an emergent phenomenon. The emphasis has been
on the $N\to\infty$ limit of the models in which noninteracting strings
form. Interactions will be present in these models at order $1/N$.
But although these interactions will be consistent with unitarity
(by construction), they will not have the complexity to reproduce
the interactions required of superstring theory.

By their very nature, $1/N$ corrections can be interpreted as
breaking and joining closed strings. This is because a
noninteracting string state is a single trace state, and
$1/N$ corrections to the Hamiltonian acting on a multi-trace
state  either split one trace into two or join
two traces into one. However as shown in
\cite{brinkgs} type IIB superstring theory requires that 
a rather complicated combination of coordinate and Grassmann
fields be inserted at the joining point of the three string
vertex. This insertion
is quadratic in coordinate fields and an 8th order
polynomial in Grassmann fields with the structure
\bea
{\cal I}_3(\sigma)&=&\tilde{X}^i(\sigma)X^j(\sigma)v^{ij}(Y(\sigma))
\eea 
In this formula $\tilde{X}$ and $X$ are linear combinations of the
worldsheet coordinate fields and $Y$ is a linear combination of the
worldsheet Grassmann fields.  The function $v^{ij}$ is a polynomial
of $Y$ constructed from the following five monomials
\bea
\delta_{ij},\qquad \gamma^{ik}_{[ab}\gamma^{jk}_{cd]}Y^aY^bY^cY^d,\qquad
\delta_{ij}Y^1Y^2\cdots Y^8,\qquad \gamma^{ij}_{ab}Y^aY^b,\qquad
\gamma^{ij}_{ab}\epsilon^{abc\cdots h}Y^c\cdots Y^h
\eea
which transform as a 2-tensor in transverse space. Here $\gamma^{ij}_{ab}
=\gamma^i_{a\dot{a}}\gamma^j_{b\dot{a}}-\delta_{ij}\delta_{ab}$,
with $\gamma^i$ the SO(8) gamma matrices.
The $1/N$ corrections from the
Hamiltonian presented in this paper can produce at most
two factors of the Grassmann fields $Y$. To produce higher
powers of $Y$, terms with more intricate
spin structure must be added to the string bit Hamiltonian. In order
to leave the large $N$ limit unaffected, we require these added
terms to have the color structure
\bea
\frac{g^{ABCD}}{N}\Tr:{\bar\phi}_A\phi_B{\bar\phi}_C\phi_D:
\eea
where the $A,B,C,D$ signify the collection of spin and flavor indices
carried by each $\phi$, and the colons indicate
normal ordering. Such terms will not contribute in leading order
in the $1/N$ expansion. An important consequence is that 
the size of the coefficient $g^{ABCD}$ can be allowed to be
very large with the perturbation still small, as long as $N\gg g^{ABCD}$.
This flexibility will particularly crucial if,as hinted by
the analysis of \cite{sunthorn}, the stability of closed chains
requires $M<N$. In that case, $N$ would have to be enormous to
produce string-like chains. In that case the only relevant interactions
would be those enhanced by a large coefficient.

It is a straightforward task to apply
such terms to multi-trace states and determine the spinor
index dependence of $g^{ABCD}$ which produces each of the monomials
listed above. The flavor dependence responsible for the $X$, $\tilde{X}$
factors of the insertion is less obvious because the emergence of
the effective coordinate fields from the Heisenberg chain dynamics is less
direct than the emergence of the Grassmann fields. 
Effective coordinate fields only arise for large bit number chains. 
The construction of a string bit Hamiltonian which implies the fully 
interacting superstring in 10 spacetime dimensions remains a project for
future research.

\vskip8pt
\noindent\underline{Acknowledgments:}  I thank Oren Bergman, John Klauder, and
Songge Sun for helpful discussions. This work is supported in part 
by U.S. DOE grant DE-FG02-97ER-41029.

\appendix
\section{Energy analysis for $-1<\Delta<1$, and large $M$}
Yang and Yang \cite{yangyang}
use the following techniques to solve the spin chain energies
in the limit $M\to\infty$. 
They map the $p_j$
onto new variables $\alpha_j$ for which $\theta$ depends only on the
difference $\alpha_j-\alpha_l$. This is accomplished by the map
\begin{eqnarray}
z&=&e^{ip}={e^{i\mu}-e^\alpha\over e^{i\mu+\alpha}-1}\nonumber\\
\Delta&=&-\cos\mu.
\end{eqnarray}
This version of the map is appropriate for $-1<\Delta<1$.
Some special values of $\alpha$ delineate the map: 
$\alpha=0$
corresponds to $e^{ip}=1$ which implies $p=0$, and $\alpha=\pm\infty$
map to $p=\pm(\pi-\mu)$. (We are choosing $p$ to be in the 
range $-\pi<p<\pi$.) Thus the whole range $-\infty<\alpha<\infty$
corresponds to $-(\pi-\mu)<p<\pi-\mu$. Note that $\Delta=1$
shrinks the range of $p$ to 0, whereas $\Delta=-1$ represents the
maximum range. It is straightforward to work out
the following quantities in terms of the new variables: 
\begin{eqnarray}
\cos p&=&-\cos\mu+\frac{\sin^2\mu}{\cosh\alpha-\cos\mu}\nonumber\\
{dp\over d\alpha}&=&\frac{\sin p}{\sinh\alpha}=\frac{\sin\mu}
{\cosh\alpha-\cos\mu}\nonumber\\
\theta(\alpha,\beta)&=&2\tan^{-1}\left[(\cot\mu)\tanh\frac{\beta-\alpha}{2}
\right]
\end{eqnarray}
The boundary conditions
take the form
\begin{eqnarray}
p_l&=&{2\pi I_l\over M}-{1\over M}\sum_{j\neq l}\theta(\alpha_j,\alpha_l),
\end{eqnarray}
where the $I_l$ are integers when $q$ is odd, and they are half-odd integers
when $q$ is even. Different choices for these integers lead
to different solutions for the set of $p$'s.  
\subsection{Consecutive $I_l$: $Q,P\neq0$}
We begin by first choosing the set of numbers $I_l$ to be consecutive
with no gaps: $I_{l+1}=1+I_l$. We define a kernel $K$ and density function
$R(\alpha)$ by
\begin{eqnarray}
K(\alpha,\beta)&\equiv&{1\over2\pi}{\partial\theta\over\partial\beta}
={1\over2\pi}{\sin2\mu\over\cosh(\alpha-\beta)-\cos2\mu}\nonumber\\
R(\alpha)&=&{2\pi\over M}{dj\over d\alpha},
\end{eqnarray}
and then convert the equation for the $p$'s as $M\to\infty$ into
an integral equation
\begin{eqnarray}
{dp\over d\alpha}&=&R(\alpha)+\int_{\alpha_-}^{\alpha_+}{d\beta}
K(\alpha-\beta)R(\beta).
\end{eqnarray}
This equation was analyzed in \cite{yangyang} for $\alpha_-=-\alpha_+$.
The values chosen for $\alpha_\pm$ determine the characteristics of the
eigenstate. For example, the eigenstate with the lowest energy
corresponds to $\alpha_\pm=\pm\infty$.
The values of $p$ at the limits of this range are $p=\pm(\pi-\mu)$.
 As long as $0<\mu<\pi$,
$e(\alpha)=4(\Delta-\cos p(\alpha))<0$ for all finite $\alpha$, 
so taking the whole range of
$\alpha$ corresponds to including in the expression for $E$
all values for $e$ less than 0. For the continuum limit
we are only interested in very large $\alpha_\pm$ since then
the eigenvalues will be close (within $1/M$) of the minimum energy
eigenvalue.

As shown in \cite{yangyang}, 
the kernel $J=-(I+K)^{-1}K$, can be used to rewrite the equation
for $R$, which determines it over the whole range of $\alpha$,
in terms of its values outside the range $(\alpha_-,\alpha_+)$.
This is useful since we are interested only in the
excited states close to the ground state corresponding to
$\alpha_\pm=\pm\infty$.
\begin{eqnarray}
R(\alpha)=R_0(\alpha)
-\left[\int_{-\infty}^{\alpha_-}+\int^{\infty}_{\alpha_+}\right]
J(\alpha-\beta)R(\beta)
\label{reqoutside}
\end{eqnarray}
where $R_0$ is the solution of the equation for $\alpha_\pm=\pm\infty$. 
It can be easily found by Fourier
transformation of the equation. From
\begin{eqnarray}
{dp\over d\alpha}&=&\int d\lambda e^{-i\lambda\alpha}
\frac{\sinh(\pi-\mu)\lambda}{\sinh\pi\lambda}\nonumber\\
K(\alpha)&=&\int_{-\infty}^{\infty}{d\lambda\over2\pi}\ e^{-i\lambda\alpha}\
{\sinh(\pi-2\mu)\lambda\over\sinh\pi\lambda},
\end{eqnarray}
we determine
\begin{eqnarray}
R_0(\alpha)&=&\int_{-\infty}^{\infty}{d\lambda}\ e^{-i\lambda\alpha}\
{1\over2\cosh\mu\lambda}\nonumber\\
&=&{\pi\over2\mu}{1\over\cosh(\pi\alpha/(2\mu))}.
\end{eqnarray} 
We can also easily express $J$ as a Fourier integral:
\begin{eqnarray}
J(\alpha)&=&-\int_{-\infty}^{\infty}{d\lambda\over2\pi}\ e^{-i\lambda\alpha}\
{\sinh(\pi-2\mu)\lambda\over2\sinh(\pi-\mu)\lambda\cosh\mu\lambda}.
\end{eqnarray}
The conserved quantities $Q=M-2q, P=\sum_jp_j$, 
the total charge and total momentum
respectively can be expressed, in the limit
$M\to\infty$, as integrals either inside or outside
the range $(\alpha_-,\alpha_+)$. These
expressions then implicitly determine $\alpha_\pm$ in terms of $Q, P$.
\begin{eqnarray}
{1\over2}-{Q\over2M}={q\over M}&=&
\int_{\alpha_-}^{\alpha_+}{d\beta\over2\pi}R(\beta)
\nonumber\\
&=&\int_{-\infty}^{\infty}{d\beta\over2\pi}R(\beta)
-\left[\int_{-\infty}^{\alpha_-}+\int^{\infty}_{\alpha_+}\right]
{d\beta\over2\pi}R(\beta)\nonumber\\
&=&{1\over2}-\left[\int_{-\infty}^{\alpha_-}+\int^{\infty}_{\alpha_+}\right]
{d\beta\over2\pi}R(\beta)\left(1+\int_{-\infty}^{\infty}d\alpha 
J(\alpha-\beta)\right)
\end{eqnarray}
Now, 
\begin{eqnarray}
1+\int_{-\infty}^{\infty}d\alpha J(\alpha-\beta)
&=&1-{\pi-2\mu\over2(\pi-\mu)}={\pi\over2(\pi-\mu)},
\end{eqnarray}
so we have
\begin{eqnarray}
{Q\over M}={\pi\over \pi-\mu}\left[\int_{-\infty}^{\alpha_-}
+\int^{\infty}_{\alpha_+}\right]{d\beta\over2\pi}R(\beta).
\end{eqnarray}
In a similar manner we can express the total momentum as
\begin{eqnarray}
{P\over M}={1\over M}\sum_{j=1}^q p_j&=&
\int_{\alpha_-}^{\alpha_+}{d\beta\over2\pi}R(\beta)p(\beta)\nonumber\\
&=&\int_{-\infty}^{\infty}{d\beta\over2\pi}R(\beta)
p(\beta)
-\left[\int_{-\infty}^{\alpha_-}+\int^{\infty}_{\alpha_+}\right]
{d\beta\over2\pi}R(\beta)p(\beta)\nonumber\\
&=&{P_0\over M}-\left[\int_{-\infty}^{\alpha_-}+\int^{\infty}_{\alpha_+}\right]
{d\beta\over2\pi}R(\beta)\left(p(\beta)
+\int_{-\infty}^\infty d\alpha J(\alpha-\beta)p(\alpha)
\right)
\end{eqnarray}
We can infer the Fourier transform of $p(\alpha)$ from that of $dp/d\alpha$.
\begin{eqnarray}
{dp\over d\alpha}&=&\int d\lambda e^{-i\lambda\alpha}{\sinh(\pi-\mu)\lambda
\over\sinh\pi\lambda}\nonumber\\
p(\beta)&=&-{1\over2i}\int d\lambda e^{-i\lambda\beta}
{\sinh(\pi-\mu)\lambda\over\sinh\pi\lambda}\left[
{1\over\lambda+i\epsilon}+{1\over\lambda-i\epsilon}\right]\nonumber\\
p(\beta)+\int d\alpha J(\alpha-\beta)p(\alpha)
&=&-{1\over2i}\int d\lambda e^{-i\lambda\beta}
{1\over2\cosh\mu\lambda}\left[
{1\over\lambda+i\epsilon}+{1\over\lambda-i\epsilon}\right]\nonumber\\
&\to& \pm{\pi\over2},\qquad {\rm for} \quad\beta\to\pm\infty.
\end{eqnarray}
Note that the $i\epsilon$ prescription is chosen so that 
$p(\pm\infty)=\pm(\pi-\mu)$, as required by the mapping.
Finally, since $P_0=0$, we have for large $\alpha_+,\alpha_-$,
\begin{eqnarray}
{P\over M}&\approx&-{\pi\over2}\left[\int^{\infty}_{\alpha_+}
-\int_{-\infty}^{\alpha_-}
\right]{d\beta\over2\pi}R(\beta).
\end{eqnarray}

Finally, we manipulate the expression for the energy,
 expressing it as an integral outside the interval 
$(\alpha_-,\alpha_+)$:
\begin{eqnarray}
\frac{E}{M}+\Delta&=&\int_{\alpha_-}^{\alpha_+}{d\beta\over2\pi}R(\beta)
e(\beta)\nonumber\\
&=&\int_{-\infty}^{\infty}{d\beta\over2\pi}R(\beta)
e(\beta)
-\left[\int_{-\infty}^{\alpha_-}+\int^{\infty}_{\alpha_+}\right]
{d\beta\over2\pi}R(\beta)e(\beta)]\nonumber\\
&=&{E_0\over M}+\Delta
-\left[\int_{-\infty}^{\alpha_-}+\int^{\infty}_{\alpha_+}\right]
{d\beta\over2\pi}R(\beta)\left(e(\beta)
+\int_{-\infty}^\infty d\alpha J(\alpha-\beta)e(\alpha)
\right),
\end{eqnarray}
where we have defined
\bea
e(\alpha)&=&4\Delta-4\cos p=\frac{-4\sin^2\mu}{\cosh\alpha-\cos\mu}=
-4\sin\mu\frac{dp}{d\alpha}\\
&=&-4\sin\mu\int d\lambda e^{-i\lambda\alpha}{\sinh(\pi-\mu)\lambda
\over\sinh\pi\lambda}
\eea
where the last line gives the Fourier transform of $e(\alpha)$.
Also $E_0$ is the energy when $\alpha_\pm\to\infty$ and $M$ is
large:
\bea
\frac{E_0}{M}+\Delta&=&\int_{-\infty}^{\infty}{d\beta\over2\pi}
R_0(\beta)e(\beta)\\
&=&-4\sin^2\mu\int_{-\infty}^{\infty}{d\beta\over2\pi}
\frac{\pi}{2\mu}\frac{1}{\cosh(\pi\beta/(2\mu))(\cosh\beta-\cos\mu)}.
\eea
This integral is not elementary for general $\mu$. But for $\Delta=0$
($\mu=\pi/2$) it is easily done;
\bea
\int_{-\infty}^{\infty}{d\beta\over2\pi}
\frac{1}{\cosh^2\beta}=\frac{1}{\pi}.
\eea
so that $E_0\to-4M/\pi$ for $\mu\to\pi/2$ or $\delta\to0$. 
Comparison to our explicit
evaluation at $\Delta=0$ shows that the correction $-4\pi/6M$ is not
included in $E_0$. That is because this term is the $M^{-1}$
correction to the replacement of the sum over discrete momenta
by an integral.

We finally arrive at a convenient expression for $E-E_0$
\begin{eqnarray}
{E-E_0\over M}&=&+4\sin\mu
\left[\int_{-\infty}^{\alpha_-}+\int^{\infty}_{\alpha_+}\right]
{d\beta\over2\pi}R(\beta)\int_{-\infty}^\infty d\lambda e^{-i\lambda\beta}
{1\over2\cosh\mu\lambda}\nonumber\\
&=&\frac{2\pi\sin\mu}{\mu}\left[\int_{-\infty}^{\alpha_-}
+\int^{\infty}_{\alpha_+}\right]
{d\beta\over2\pi}R(\beta){1\over\cosh(\pi\beta/2\mu)},
\end{eqnarray}
To find the energy levels close to the ground state, we must 
analyze the equations for $R$ for large $\alpha_+, \alpha_-$.
For $\alpha>\alpha_+$, (\ref{reqoutside}) can be approximated
by dropping the integral over negative $\alpha$ and using the
asymptotic form for $R_0$:
\begin{eqnarray}
R(\alpha)+\int^{\infty}_{\alpha_+}J(\alpha-\beta)R(\beta)&\approx&
{\pi\over\mu}e^{-\pi\alpha/2\mu}.
\label{req+}
\end{eqnarray}
It is convenient to put 
$$R(\alpha+\alpha^+)
={\pi\over\mu}e^{-\pi\alpha_+/2\mu}S(\alpha)$$
so that (\ref{req+}) reduces to the Wiener-Hopf equation \cite{yangyang}
\begin{eqnarray}
S(\alpha)+\int^{\infty}_0 J(\alpha-\beta)S(\beta)&=&
e^{-\pi\alpha/2\mu}.
\label{wienerhopf}
\end{eqnarray}
Similarly, analyzing the equation for $\alpha<\alpha_-$, leads to
the identification
$$R(\alpha+\alpha^-)
\approx{\pi\over\mu}e^{\pi\alpha_-/2\mu}S(-\alpha).$$
Inserting these approximations into the formulas for $Q$, $P$,
and $E$, leads to 
\begin{eqnarray}
{Q\over M}&\approx&{\pi\over\pi-\mu}{1\over2\mu}
\left[e^{-\pi\alpha_+/2\mu}+e^{\pi\alpha_-/2\mu}\right]\int_0^\infty
d\beta S(\beta)\nonumber\\
{P\over M}&\approx&-{\pi\over2}{1\over2\mu}
\left[e^{-\pi\alpha_+/2\mu}-e^{\pi\alpha_-/2\mu}\right]\int_0^\infty
d\beta S(\beta)\nonumber\\
{E-E_0\over M}&\approx&\frac{2\pi\sin\mu}{\mu} \frac{1}{\mu}
\left[e^{-\pi\alpha_+/\mu}+e^{\pi\alpha_-/\mu}\right]\int_0^\infty
d\beta S(\beta)e^{-\pi\beta/2\mu}.
\end{eqnarray}
Next one can solve the first two equations for $\alpha_+$ and
$\alpha_-$ and substitute in the last equation to get
\begin{eqnarray}
{E-E_0\over M}&\approx&\frac{2\pi\sin\mu}{\mu}
{2\mu I(\pi/2\mu)\over I(0)^2}
\left[{(\pi-\mu)^2\over\pi^2}{Q^2\over M^2}+{4\over\pi^2}
{P^2\over M^2}\right],
\end{eqnarray}
where we have defined $I(x)=\int_0^\infty d\beta S(\beta)e^{-x\beta}$.
From the solution of (\ref{wienerhopf}), one can infer (see
\cite{yangyang}) that 
${I(\pi/2\mu)/I(0)^2}=\pi^2/8\mu(\pi-\mu)$, so finally
\begin{eqnarray}
{E-E_0}&\approx&\frac{2\pi\sin\mu}{\mu}{1\over M}
\left[{\pi-\mu\over4}{Q^2}+{1\over\pi-\mu}{P^2}\right]
\label{gaplessexc}
\end{eqnarray}

\subsection{Non-consecutive $I_l$}
The excited states included
in (\ref{gaplessexc}) are those where the numbers $I_l$
are consecutive. For example, the state with $Q=P=0$
corresponds to the choice (with $q=M/2$ odd)
$$\left(-{q-1\over2},\ldots,{q-3\over2},{q-1\over2}\right).$$
There are also excitations in which
``holes'' are allowed in this set of numbers. As an example,
consider replacing $(q-1-2j)/2$ in the above list by $(q+1)/2$,
creating a gap, but retaining the same number
of overturned arrows, so that $Q=0$. However the momentum
is increased by the amount $P=2\pi(j+1)/M$. For large $M$,
the effect of this hole on the $p$'s is small, and it
makes sense to expand them around the values appropriate
to the $Q=P=0$ state, the new set of $p$'s differing from the
latter by $\delta p_j$. Referring to the original equation
for the $p$'s, we find an equation for $\delta p$:
\begin{eqnarray}
\delta p_l&=&{2\pi\over M}\theta(l-l_j)+{2\pi \over M}
\sum_{j\neq l}\left[-{\partial\alpha_j\over\partial p_j}
\delta p_j+{\partial\alpha_l\over\partial p_l}
\delta p_l\right]K(\alpha_l-\alpha_j)\nonumber\\
\delta p_l\left(1-{\partial\alpha_l\over\partial p_l}
{2\pi\over M}\sum_{j\neq l}K(\alpha_l-\alpha_j)\right)
&=&{2\pi\over M}\theta(l-l_j)-{2\pi\over M}
\sum_{j\neq l}\left[{\partial\alpha_j\over\partial p_j}
\delta p_l\right]K(\alpha_l-\alpha_j)\nonumber\\
\delta p(\alpha){\partial\alpha\over\partial p}R(\alpha)
&=&{2\pi\over M}\theta(\alpha-\alpha_j)-
\int d\beta{\partial\beta\over\partial p}
\delta p(\beta)R(\beta)K(\alpha-\beta),
\end{eqnarray}
where we have replaced the sums by integrals in the last line.
Defining 
$$\chi(\alpha)=M\delta p(\alpha)R(\alpha)d\alpha/dp,$$
we have the integral equation
\begin{eqnarray}
\chi(\alpha)+\int_{-\infty}^\infty d\beta K(\alpha-\beta)\chi(\beta)
&=&2\pi\theta(\alpha-\alpha_j).
\label{pheq}
\end{eqnarray}
Here $\alpha_j$ marks the location of the ``hole''. It can be 
related to the value for the momentum of the excited state:
\begin{eqnarray}
P&=&{2\pi(j+1)\over M}=\sum_l \delta p_l
=\int_{-\infty}^\infty {d\alpha\over2\pi}\chi(\alpha)
{dp\over d\alpha}
\end{eqnarray}
Similarly, we can write the energy difference between the excited
and ground state as
\begin{eqnarray}
E-E_0&=&\sum_l\delta p_l
{d\alpha_l\over dp_l}{de(\alpha_l)\over d\alpha_l}\nonumber\\
&=&-4\sin\mu\int_{-\infty}^\infty {d\alpha\over2\pi}\chi(\alpha)
\left[\int d\lambda e^{-i\lambda\alpha}{-i\lambda\sinh(\pi-\mu)\lambda
\over\sinh\pi\lambda}\right]\nonumber\\
\end{eqnarray}
(\ref{pheq}) can be immediately solved via Fourier transformation:
\begin{eqnarray}
\chi(\alpha)&=&i\int d\lambda e^{-i(\alpha-\alpha_j)\lambda}{\sinh\pi\lambda
\over2(\lambda+i\epsilon)\sinh(\pi-\mu)\lambda \cosh\mu\lambda},
\end{eqnarray}
and used to obtain the total momentum and energy 
\begin{eqnarray}
P&=&{i\over2}\int{d\lambda}\ {e^{-i\lambda\alpha_j}
}{1\over(-\lambda+i\epsilon)\cosh\mu\lambda}\nonumber\\
E-E_0&=&2\sin\mu\int{d\lambda}\ {e^{-i\lambda\alpha_j}
}{1\over\cosh\mu\lambda}.
\end{eqnarray}
Of course, we are interested in these expressions in the limit
$\alpha_j\to\infty$, corresponding to the continuum limit. This
asymptotic limit is obtained by deforming the integration 
contours into the lower half plane and picking up the
nearest pole to the real axis, namely the one at $\lambda=-i\pi/2\mu$.
This leads to
\begin{eqnarray}
P\sim 2 e^{-\pi\alpha_j/2\mu} \qquad\qquad E-E_0\sim
2\frac{2\pi\sin\mu}{\mu}
e^{-\pi\alpha_j/2\mu},
\end{eqnarray}
from which we conclude that
\begin{eqnarray}
E-E_0=\frac{2\pi\sin\mu}{\mu}{P}=\frac{2\pi\sin\mu}{\mu}{2\pi(j+1)\over M}
\end{eqnarray}
in the limit $M\to\infty$. Notice the important fact that the
energy of these excitations is the factor $(2\pi\sin\mu)/\mu$,
common to the other contributions to $E$,
times a coefficient independent of $\mu$. Although we 
have discussed only one particular ``particle-hole''
excitation, it is clear that the energy of the state with many particle-hole
pairs will simply be additive in the momentum carried by each
pair. Furthermore, there are two independent
sets of such excitations about the two boundaries of the
Fermi sea. Each particle hole excitation contributes $2\pi nT_0/P^+$,
where $n>0$. If there are several particle-hole pairs from the right
side  $p>0$ of the Fermi sea, we define $N_R=\sum_i n_i$, and similarly $N_L$
is defined for those from the left side $p<0$ of the Fermi sea. 

These contributions to the energy are added to those arising
from non-zero $Q, P$.
Note that the $P^2$ term in the energy receives negligible contributions
from particle-hole excitations from the same side of the Fermi
sea, since these have $P=O(1/M)$. 
This term is non-zero in the continuum limit only if the particle and hole
are from opposite sides of the sea. For example, replacing
$-(q+1)/2$ with $(q+1)/2$ contributes $2\pi q/M\approx\pi$ to
$P$. But such large momentum pair excitations have already been accounted
for among the excitations with consecutive $I_l$ considered earlier.
Thus the energy levels of the continuum limit are determined
by $Q$, $P$, $N_R$, and $N_L$:
\begin{eqnarray}
E-E_0&=&\frac{2\pi\sin\mu}{\mu}
\left[-\frac{\pi}{6}+{\pi-\mu\over4}{Q^2}+{1\over\pi-\mu}{P^2}
+{2\pi(N_R+N_L)}\right]\frac{1}{M}.
\label{ferroen2}
\end{eqnarray}
Recall that $Q=2r$ and $P=\pi s$ where $r, s$ range independently over all
integers. Here we have taken the liberty of inserting the $-\pi/(6M)$
correction, whose value should be the same relative to 
$2\pi(N_R+N_L)$ as in the $\Delta=0$ case. This is because this
term is just the zero point energy associated with the particle hole
excitations \cite{brinknielsen}.

\end{document}